\documentclass[twocolumn,showpacs,preprintnumbers,amsmath,amssymb]{revtex4}
 \usepackage{graphicx}
 \usepackage{dcolumn}
 \usepackage{bm}
 
\begin{document}

 \title{Renormalization of f-levels away from the Fermi energy in electron
 excitation spectroscopies: Density functional results of
 Nd$_{2-x}$Ce$_x$CuO$_4$}

 \author{T. Jarlborg$^1$, B. Barbiellini$^2$, H. Lin$^2$, R.S.
 Markiewicz$^2$, A. Bansil$^2$.}

 \affiliation{
 $^1$DPMC, University of Geneva, 24 Quai Ernest-Ansermet, CH-1211 Geneva 4,
 Switzerland
 \\
 $^2$Department of Physics, Northeastern University, Boston, Massachusetts
 02115, USA
 }
 
 
 \begin{abstract}
 Relaxation energies for photoemission where an occupied electronic state is
 excited and for inverse photoemission where an empty state is filled are
 calculated within the density
 functional theory with application to Nd$_{2-x}$Ce$_x$CuO$_4$. The
 associated relaxation energies are obtained by computing differences in
 total energies between the ground state and an excited state in which one
 hole or one electron is added into the system. The relaxation energies
 of f-electrons are found to be of the order of several eV's, indicating
 that f-bands will appear substantially away from the Fermi energy ($E_F$)
 in their spectroscopic images, even if these bands lie near $E_F$.
 Similar shifts are obtained for the Gd-f states in Gd$_2$CuO$_4$.
 Our analysis explains why it would be difficult
 to observe f electrons at the $E_F$ even in the absence of strong  electronic
 correlations.
 \end{abstract}
 
 
 \maketitle
 
 \section{Introduction}
 Compounds containing rare-earth (RE) or actinide elements display many
 intriguing solid state phenomena such as heavy fermion behavior and
 high-temperature superconductivity. When partially filled f-orbitals are
 involved, the ground state predicted by the density functional theory (DFT)
 clearly places the f-electrons in narrow bands piled at the Fermi energy $E_F$,
 interacting only weakly with other electrons. In sharp contrast however
 signatures of f-bands are often found in spectroscopic measurements not at
 $E_F$, as band theory predicts, but several eVs above or below the $E_F$
 depending on the nature of the spectroscopy \cite{lang,marel,hufner}. 
 Here we show how
 this dilemma can be resolved and how this seemingly contradictory behavior of
 f-electrons in solid-state systems can be modeled within the framework of the
 DFT. In essence, we have carried out DFT calculations constrained to simulate
 the process of electron excitation. In this way, we adduce that when f-bands are
 excited, their spectroscopic image will generally avoid $E_F$ even when these
 bands lie at the $E_F$.

 We consider NCCO as an exemplar complex system, which is interesting not only
 because it is a high-$T_c$ superconductor \cite{armi}, but also because it
 contains Nd-f bands co-existing with the broader Cu-O bands \cite{fulde}. This
 allows us to delineate relative differences in the way localized bands get
 excited in comparison to the itinerant bands. Our computations are based on the
 final state rule and the $\Delta$-self-consistent field ($\Delta$SCF) method,
 which have been invoked previously to investigate x-ray absorption and emission
 \cite{bart,wein,lerc}, and x-ray photoemission spectroscopy (XPS) of core levels
 \cite{taka}. The final state rule postulates that the electronic structure of
 the excited state can be obtained by using the potential of the final state in
 which the hole is present in the valence band, i.e. the hole is long-lived and
 the electronic system relaxes before the hole recombines with an electron. In
 standard computations, the excitation energies are approximated by band
 energies, but in the $\Delta$SCF method the energies are significantly more
 realistic because they are obtained by computing the total energy difference
 between the unperturbed ground state and a relaxed cell calculation for the
 final state. Our goal is to capture the fundamental mechanism responsible for
 why f-electron excitations avoid $E_F$, without consideration of probe dependent
 matrix elements \cite{foot1,newarpes,rixs,stm,compton,positron}. For this
 purpose, we introduce a method beyond the many body perturbation theory
 \cite{bba05, tiago08, dabo10, rohlfing00} for calculating the energy cost of
 localized f-state relaxations.

 \section{Method of calculation}
 Filled states can be probed via x-ray emission or photoemission spectroscopy,
 and empty states through x-ray absorption or inverse photoemission or
 bremsstrahlung isochromat spectroscopy, issues of surface sensitivity of
 photoemission notwithstanding. Formally, the energy for exciting an electron
 requires us to compute the difference between the total energy $E_0(N)$ of the
 $N$-particle ground state and the energy $E_n(N-1)$ of an excited state of
 $(N-1)$ particles containing a hole in the $n^{th}$ level \cite{dyson}. The
 relaxed excitation energy $\epsilon_n$ of the $n^{th}$ filled level then is
 \begin{equation}
 \epsilon_n = E_n(N-1) - E_0(N).
 \label{eq_excite}
 \end{equation}
 The energy contribution from the photon, $\hbar\omega$, can be subtracted off,
 since it is not an interesting part of the relaxation energy. Similarly, for the
 inverse process, where we add an electron into an empty state of the
 $N$-particle system, we need to evaluate the total energy $E_n(N+1)$ of an
 excited state of the $(N+1)$ particle system, so that the excitation energy
 $\epsilon_n$ of the n$^{th}$ unfilled state is
 \begin{equation}
 \epsilon_n = E_n(N+1) - E_0(N).
 \label{eq_iexcite}
 \end{equation}
 In the absence of relaxation, $\epsilon_n = \epsilon_n^{0}$,
 where $\epsilon_n^{0}$ is the energy of the $n^{th}$ Kohn-Sham
 orbital \cite{lda}. The correction to this Kohn-Sham energy is the relaxation
 energy
 \begin{equation}
 E_r^{(n)} = \epsilon_n - \epsilon_n^{0} \label{eq_relax}
 \end{equation}
 associated with the excitation process.
 The connection of groundstate Kohn-Sham equations to one particle energies
 has been studied by Bauer \cite{bauer83}.
 Other authors \cite{duffy94,bba05} have shown
 that the exchange-correlation potential $V_{xc}$ for
 calculating ground state properties is the best local approximation to the
 exchange-correlation self-energy at the Fermi level in Dyson's quasiparticle
 equation.

\begin{figure}[h]
 \begin{center}
 \includegraphics[width=3.01in]{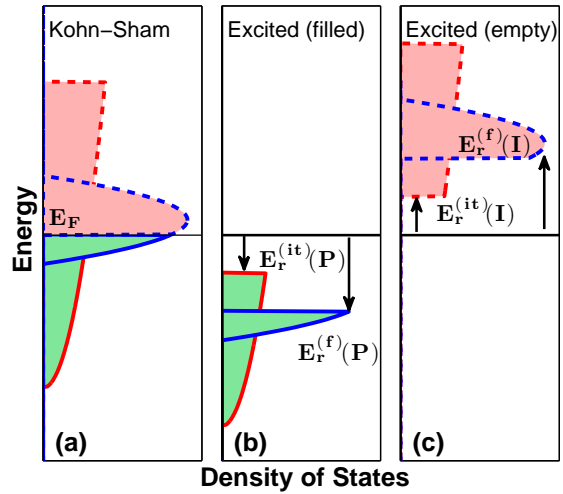}
 \end{center}
 \caption{(Color online)
 Schematic illustration of how the Kohn-Sham spectrum is mapped in
 excitation processes in the presence of relaxation effects. (a) Kohn-Sham DOS
 in which the filled portion is shaded green and the empty portion is
 shaded pink. The filled f-band is shown by a solid blue line and the itinerant band
 by a solid red line. The unfilled portions of these bands are marked by
 broken lines of the same color. $E_F$ is the Kohn-Sham Fermi energy.
 (b) Image of the filled portion of the DOS where f and itinerant bands
 are shifted by $E_r^{(f)}(P)$ and $E_r^{(it)}(P)$, respectively, in
 exciting an electron. (c) Same as panel (b), except that here the unfilled
 portion of the DOS is excited and the f and itinerant bands undergo
 relaxation shifts of $E_r^{(f)}(I)$ and $E_r^{(it)}(I)$.
 }
 \label{fig1}
 \end{figure}

 Fig.~\ref{fig1} illustrates how a partially filled f-band lying at the $E_F$,
 which is superposed on a broader band, will be mapped in the excitation process.
 If the relaxation energy $E_r^{(n)}$ is zero or a
 constant, then the spectroscopic image of the Kohn-Sham density of states (DOS)
 in Fig.~\ref{fig1}(a) obtained, for example, via photoemission and inverse
 photoemission processes, will be an undistorted copy of the DOS \cite{foot1}. In
 general, however, $E_r^{(n)}$ will be non-zero and differ between localized and
 itinerant bands, and between occupied and empty states. Therefore, it is useful
 to introduce the notation $E_r^{(f)}(P)$ and $E_r^{(it)}(P)$ for the relaxation
 energies of the f and itinerant bands for the occupied states, which could be
 probed via photoemission. Similarly, $E_r^{(f)}(I)$ and $E_r^{(it)}(I)$ denote
 the corresponding relaxation energies for the unfilled states, which could be
 accessed via an inverse photoemission process. Fig. \ref{fig1}(b) shows that
 when the filled portion of the Kohn-Sham DOS is mapped, f-bands are shifted by
 $E_r^{(f)}(P)$ and itinerant bands by $E_r^{(it)}(P)$, so that the f-bands no
 longer appear to be at the $E_F$. Fig. \ref{fig1}(c) shows what happens when the
 unoccupied portion of the DOS is excited. Now the f-bands move by $E_r^{(f)}(I)$
 and the itinerant bands by $E_r^{(it)}(I)$. The net result is that filled and
 unfilled portions of the DOS, as seen by comparing panels (b) and (c), will in
 general appear to be separated in energy in their spectroscopic images. The
 preceding effects arise purely from the way the excitation processes play out
 and reflect differences in the screening of the added hole or electron in
 various orbitals of the unperturbed system.

 Concerning technical details, we note first that the $\Delta$SCF scheme works
 well for localized core states with pure $\ell$-character \cite{lerc}. Here, we
 apply the scheme to a localized RE f-band in NCCO, where our band calculations
 based on the local spin density approximation (LSDA) \cite{lda,lsda} and the
 Linear Muffin-Tin Orbital (LMTO) method \cite{oka,bdj,bbtj}, put the Nd-f band
 right at $E_F$. LSDA yields the correct CuO$_2$-plane Fermi surface and
 reasonable band dispersions in doped LSCO \cite{newarpes}. Also, LSDA correctly
 predicts both metallic and ferromagnetic phases in manganites \cite{bba_cmr}.
 That LSDA fails to describe the anti-ferromagnetic (AFM) phase within the
 CuO$_2$ plane in undoped cuprates is not relevant here since our focus is on the
 RE physics. Notably, however, LSDA calculations for LSCO using lower
 linearization energies describe the undoped as well as the doped
 system reasonably\cite{15}, indicating that only small corrections are needed to the LSDA
 to bring the paramagnetic and the antiferromagnetic states into the correct
 order. For undoped NCCO, LSDA does predict a metallic RE AFM ground state with
 RE order in line with experimental findings \cite{armi}.  A RE ferromagnetic
 metallic solution is very close in energy, and it is more convenient to use for
 calculating excitation energies. Accordingly, we will study excitations from the
 fictitious ferromagnetic ground state for $x =0 $, which are very similar to the
 excitations of the metallic compound with $x \neq 0$ \cite{foot2,new1,new2}. Our
 basis set includes $\ell\le3$ for Nd, $\ell\le 2$ for Cu and O, and $\ell\le1$
 for empty spheres, two of which are inserted per formula unit in the most open
 part of the structure. The lattice constant, $a_0$, is $7.45$ a.u., and the
 Wigner-Seitz radii are (in units of $a_0$): $0.44$, $0.33$, $0.31$ and $0.272$
 for Nd, Cu, O and empty spheres, respectively. The number of irreducible
 k-points used is $80$ for a cell containing four formula units. The computed
 moment on each Nd is $3.15$ $\mu_B$ while the moments on Cu and O atoms are less
 than $0.1~\mu_B$. Interestingly, by imposing the observed AFM order on Cu
 through a staggered field, the magnetic order on the RE does not change,
 implying weak interaction between Cu and RE sites. Turning to excited state
 simulations, we consider first the process in which an occupied electron is
 excited leaving behind a hole in the system. We model this hole as a hole in the
 local density of states (LDOS) on a particular site (e.g. Nd, Cu or O).
 Specifically, the hole is created by removing electrons from the LDOS over an
 energy window $[E_c,E_F]$, where $E_c \le E_F $ is a cut-off energy defined such
 that a total of one electron is removed. The total charge for the excited state
 at the site $t'$ then is
 \begin{eqnarray}
 \rho(r)=\sum_{t,\ell}\int_{-\infty}^{E_F} N_{t,\ell}(E)R^2_{t,\ell}(E,r)dE
 \nonumber \\
 -  \sum_{\ell}\int_{E_c}^{E_F}N_{t',\ell}(E)R^2_{t',\ell}(E,r)dE +
 \frac{1}{\Omega},
 \label{rho}
 \end{eqnarray}
 where $N_{t,\ell}(E)$ is the LDOS and $R_{t,\ell}(E,r)$ is the radial wave
 function at the site $t$ with angular momentum $\ell$. The last term in
 Eq.~\ref{rho} imposes charge neutrality within the simulation cell of
 volume $\Omega$. This form is appropriate for XPS, since at large
 excitation energies the wave function of the electron is free-electron
 like with a constant charge density \cite{jn}. At lower photon energies
 (e.g. for UV photoemission \cite{ups}), $1/\Omega$ could be replaced by
 the proper density for the excited state. Calculations involving core levels
 show good convergence already for
 small unit cells \cite{lerc}. The reason is that the charge screening
 occurs near the excited site. We have embedded one excited site
 in a fairly large cell of $28$ sites, and  verified that both charges
 and local DOS on sites far from the excitation approach those of the
 ground state for all cases.
 The constraint of Eq.~\ref{rho} is repeated for each
 self-consistent iteration until the total energy is converged to an
 accuracy of about $0.01$ eV. The excitation energy $\epsilon_n$ is calculated
 by a change in total energy between the ground state and the excited state as
 in Eq.~\ref{eq_excite}.
%
%

 \section{Results}
 Before discussing our results for the relaxation energies with reference
 to Table 1, we emphasize that in our modeling the hole does not
 involve a single energy level of the solid, but rather a group of states
 around the mean energy $\bar\epsilon(P)$ \cite{footnebar}.
 The values of
 $\bar\epsilon (P)$ are given in the second column of the Table.
 $\bar\epsilon (P)$ is quite small ($0.04$ eV) for the majority (up spin) Nd
 band, dominated by f-levels, so that this hole state is modeled reasonably in
 our scheme \cite{footscheme}. Values of $\bar\epsilon (P)$ for other states
 considered (minority spin Nd, Cu,
 and apical and planar O atoms) range from $1-4$ eV, and thus the hole in
 these cases involves a substantial mixture of states around the mean
 value.
 
 Focusing on the relaxation energies $E_r^{(n)}(P)$ of the occupied states
 given in the 3rd column of the Table, we see that the up spin
 Nd state, which is almost exclusively of f-character undergoes a relaxation
 shift of $-4.45$ eV, so that in the excitation
 spectrum this state will appear at an excitation energy of $-0.04-4.45=-4.49$ eV,
 i.e. at a much lower energy than in the Kohn-Sham spectrum. In
 contrast, an excitation from the down spin Nd band, which contains almost no
 f-electron (but involves itinerant d-electrons), with $\bar{\epsilon} = -3.63$
 eV will appear at $-0.68$ eV, i.e. at a much higher energy than for the up
 spin band.
 Along these lines, Cu excitations experience a
 relaxation correction of about $-2$ eV, while the corrections for
 planar and non-planar O-atoms [O(1) and O(2)] are both
 about $+0.6$ eV. As expected,
 the more localized levels generally suffer larger relaxation effects. In
 particular, the majority spin Nd bands of f-character, which are most strongly
 localized,
 display the largest correction. The Cu-d band is also relatively localized
 and undergoes substantial hole screening. But Cu and O states are
 hybridized so that actual excitations from the itinerant CuO bands will be
 an average over the Cu and O contributions in the Table. [This is not the
 case for Nd-f bands, which hardly hybridize with other valence states.]
 Putting all this together, we estimate that an excitation from the
 majority Nd-f will be detected $\sim 2.5$  eV lower than from other bands
 (i.e. the value of $4.45$ eV in Table 1 is reduced by about $2$ eV as the
 Fermi level follows the CuO hybrid band),
 giving the appearance of a gap between the Nd-f states and the Fermi
 level, as shown schematically in Fig.~\ref{fig1}(b). The aforementioned
 computed downward shift of about $2.5$ eV is in reasonable accord with
 photoemission measurements, which find a broad peak centered at about $-3$
 eV below $E_F$ for the 4f level \cite{klau}.
 \begin{table}[ht]
 \caption{\label{table1}
 Relaxation energies in NCCO for excitations from majority- and
 minority-spin Nd [Nd $(\uparrow)$ and Nd$(\downarrow)$], Cu, and planar and
 apical O sites [O(1) and O(2)]. $E^{(n)}_{r}(P)$ is the relaxation energy
 of photoemission with the hole at an average energy
 of $\bar{\epsilon}(P)$ and an excited electron above $E_F$. For the inverse
 process,
 the relaxation energy and the average energy of the added electron
 in the final state are  $E^{(n)}_{r}(I)$ and $\bar{\epsilon}(I)$,
 respectively. All energies are in eV.
 }
 \vskip 2mm
 \begin{center}
 \begin{tabular}{l c c c c}
 \hline
 Band & $\bar{\epsilon}(P) $ & $E^{(n)}_{r}(P)$ &  $\bar{\epsilon}(I)$  &
 $E^{(n)}_{r}(I)$ \\
 \hline \hline
 
 Nd ($\uparrow$)   &  -0.04 & -4.45 & 0.04 & 4.81  \\
 Nd ($\downarrow$) &  -3.63 & 2.95 & 1.78 & 4.38  \\
 Cu                &  -1.03 & -2.09 & 4.54 & 2.46 \\
 O(1)              &  -1.81 & 0.58 & 5.48 & 3.40  \\
 O(2)              &  -2.96 & 0.65 & 5.91 & 2.99  \\
 
 \hline
 \end{tabular}
 \end{center}
 \end{table}

 Relaxation effects in the inverse excitation process are considered in the
 last two columns of Table 1. In this case, the final state involves an
 extra electron with an average energy of $\bar{\epsilon}(I)$ and the
 corresponding relaxation energy is $E^{(n)}_{r}(I)$. Computation of the
 relaxation energy follows along the lines described above for the
 case of a final state with an extra hole, except that here we add an
 electron near the $E_F$ in the empty LDOS on various atomic sites, 
 and $E_c \ge E_F$. In
 order to ensure charge neutrality, Eq.~\ref{rho} is now modified so that the
 charge density term $1/\Omega$ is subtracted (instead of being added) on
 the right hand side of Eq.~\ref{rho}. Turning to the results for the majority
 Nd (f mainly)
 states in Table 1, we find our key result: There is a large positive shift
 $E_{r}^{(n)}(I)$ of $4.81$ eV, implying that Nd-f$(\uparrow)$ bands will
 appear to lie at energies well above $E_F$ and above the itinerant CuO bands.
 In contrast to the case of photoemission, the relaxation energies of
 Nd-$\uparrow$
 and Nd-$\downarrow$ are very similar. This is because both up- and down-spin
 f-states exist
 in the unoccupied Nd DOS, while the occupied Nd DOS has almost no down-spin
 f-electrons.
  
The Gd is a prototypical member of the rare earths 
which is a suitable benchmark for testing excitations of 4f
electron systems in X-ray photoemission (XPS) and 
bremsstrahlung isochromat (BIS) \cite{chantis}. 
In our ground state calculation for the FM configuration
of Gd$_8$Cu$_4$O$_{16}$, the majority Gd-f band is completely
filled and at about $4.5$  eV below $E_F$ while the minority Gd-f
is empty and at about $0.5$ eV  above $E_F$.  
When we apply our corrections the estimated XPS peak is at
about $-7$ eV whereas the BIS peak is near $4$ eV, which
compare fairly well with the corresponding experimental 
values in pure Gd \cite{lang}, about $-8$ and $4.5$
eV respectively.

 \section{Discussion}
 It is interesting to note that in the case of a Hubbard band strong
 on-site Coulomb replusion $U$ between electrons of opposite spin splits
 the band into a lower portion lying $\sim U/2$ below the $E_F$, and an
 upper portion lying $\sim U/2$ above the $E_F$. The gap between the upper
 and lower Hubbard bands so created is the result of strong electronic
 correlations. In contrast, in the present LSDA calculation the
 partly filled f-band is located at $E_F$, but
 screening effects in the excitation process make the filled portion appear
 well below the $E_F$ and the unfilled portion appear well above the $E_F$.
 The effective splitting between these two portions from Table 1 (first row),
 $E^{(n)}_r(I) - E^{(n)}_r(P)$, is $\sim 9$~eV for Nd. The itinerant bands in
 Table 1 are seen to display
 smaller splittings. Apparently, the relaxation energy can be sizable even for
 itinerant valence
 electrons \cite{kowalczyk75}.
 Experimental determinations of effective $U$ for elemental Nd are in
 the range 6-7 eV \cite{lang,marel}. Constrained LSDA calculations have
 been performed to estimate Hubbard $U$ parameters
 \cite{hybertsen}, and have been used to justify opening of Mott gaps in the
 cuprates.
 Similar methods have been applied for calculation of $U$ parameters for d-band
 impurities
 via relaxation energies of embedded atoms  \cite{solo}. Our computational scheme
 is however tailored for treating the process of electron or hole excitation for
 a specific pair of initial and final states. Therefore, our relaxation energies
 cannot be described in terms of a single atomic parameter, although we would
 expect our relaxation energies to be of the order of the commonly used $U$
 parameters. Our approach is in the spirit of the early
 work of Herbst and Wilkins \cite{herb} although Ref.~\onlinecite{herb} considers
 excitations from renormalized atoms within
 truncated Wigner-Seitz spheres. We also emphasize that our approach is quite
 different from LDA+U type calculations,
 since we employ the constrained DFT formalism and not the
 orbital dependent techniques invoked in LDA+U \cite{lda+u}.
 Finally, we note that our LSDA calculations do not address Mott-physics
 involving strong correlations,
 but demonstrate that final state corrections can give the appearance of a gap in
 spectroscopic data even in the absence of strong electron correlations.

 \section{Conclusion} 
 We have presented a DFT-based scheme for obtaining
 relaxation energies relevant for excitation of occupied and empty states
 in various spectroscopic probes. Our method is particularly suitable for
 f-bands and demonstrates clearly that the screening of excitations in
 f-bands differs substantially from that of more delocalized states. The
 net effect is that in the excitation spectrum the filled f-bands appear to
 move below the $E_F$ while the unfilled f-bands are shifted above the
 $E_F$. The most important
 conclusion of our study is that the aforementioned shifts are induced via
 screening that occurs within the excitation process, even when the
 majority spin f-states in NCCO lie at the $E_F$ in the LSDA Kohn-Sham
 spectrum.
 We expect this screening mechanism to be applicable to f-electron systems
 more generally since many f-electron compounds contain narrow, partially
 filled f-bands at the Kohn-Sham $E_F$ with little hybridization
 with other bands. Although we have assumed a high excitation energy for
 the generic purposes of this study, it will be straightforward to extend
 our scheme to consider lower energy excitations. For a realistic description of
 the spectral {\it intensities} in various spectroscopies, one will need to take
 the matrix element effects into account. Furthermore,
 it will be interesting to examine the extent to which f-electrons in the
 ground state can contribute to Fermi surface related properties.
 In particular, if NCCO has f-electrons at or near the $E_F$, which probes
 could detect them? Compton scattering \cite{compton1} and
 positron-annihilation \cite{positron1,shukla},
 which are sensitive to the electron momentum density of
 the many-body ground state could be promising in this connection.

 \begin{acknowledgments}
 We acknowledge useful discussions with G. Zwicknagl and P. Fulde. This
 work is supported by the US Department of Energy, Office of Science, Basic
 Energy Sciences contracts DE-FG02-07ER46352 and DE-FG02-08ER46540 (CMSN),
 and benefited from allocation of computer time at the NERSC and NU-ASCC
 computation centers. 
 \end{acknowledgments}


\begin{thebibliography}{10}
\bibitem{foot1} Matrix element effects can be quite strong and generally differ
 greatly between different highly resolved spectroscopies such as angle-resolved
 photoemission (ARPES) \cite{newarpes}, resonant inelastic x-ray scattering
 (RIXS) \cite{rixs}, scanning tunneling microscopy (STM) \cite{stm}, Compton
 scattering \cite{compton} and positron-annihilation \cite{positron}.
\bibitem{foot2} It will be interesting to examine doping effects by using
 supercell or other first principles approaches \cite{new1,new2}.
\bibitem{ups} Signatures of f-states in ultraviolet photoemission spectroscopy
 might be weak when final d-states (required for dipole transitions) are absent
 while in XPS there are always some states available for
 $\Delta\ell \pm 1$ transitions.
\bibitem{footnebar} This energy is a replacement
 of Kohn-Sham eigenvalue
 $\epsilon_n^{0}$ which is given by $\bar\epsilon(P)
 = \sum_{\ell}\int_{E_c}^{E_F} N_{t',\ell}(E)EdE -E_F$.
\bibitem{footscheme}
 Our analysis indicates that f-electron excitations are quite atomic-like
 and are almost fully contained within a Nd-sphere with little mixing with
 other orbitals. For less localized orbitals accurate estimate of
 relaxation energy should involve a generalized model of the hole state
 which allows for mixing of different orbital characters in the relaxation
 process.
\end{thebibliography}
 \end{document}